\documentclass[twocolumn]{article}
\usepackage{amsmath, amssymb}
\usepackage{mathrsfs}
\usepackage{cite}
\usepackage[dvipdfmx]{graphicx, xcolor}
\usepackage{array, booktabs}
\begin{document}
\title{\large \bf Gauge mediation without gravitino problem\\ by virtue of QCD anomaly}
\author{\normalsize Nobuki Yoshimatsu}
\date{\it \footnotesize Kawaijuku Manavis Horidome\\1-2-23\ Horidome Nishi, Wakayama-shi, Wakayama 641-0045, Japan\\ 
{\footnotesize E-mail address:\ nyoshimatsu260@gmail.com}}
\twocolumn[
\maketitle
\begin{center}
Abstract
\end{center}
{\small We consider the possibility that the Peccei--Quinn charged dynamical supersymmetry breaking (DSB) mechanism readily accommodates gauge mediation without the gravitino problem. Given a gravitino of $m_{3/2}=\mathcal{O}(eV)$ for the Intriligator--Seiberg--Shih model, an axion, converted from a Nambu--Goldstone boson, could become a dark matter candidate, while the QCD anomaly and a sizable $\mu$--value are naturally generated. We also stress that rather minimal DSB content is sufficient to yield the observed mass of the lightest Higgs boson, other than the acceptable soft mass of sparticles in the observable sector.}
{\flushleft{\ }}]

{\it Introduction}

Gauge-mediated supersymmetry (SUSY) breaking \cite{giudice rattazzi} has been recognized as a promising mechanism that plausibly communicates with the minimal supersymmetric standard model (MSSM) sector, mainly because the gauginos and sfermions acquire the flavor universal soft mass, leading to the highly suppressed flavor changing process. However, the gravitino problem might be unavoidably encountered; the relic density of the stable gravitino with $1$ keV $<m_{3/2}<1$ GeV implies a low reheating temperature, i.e., $T_R \lesssim 10^6$ GeV, which seriously conflicts with the thermal leptogenesis \cite{Fukugita Yanagida, FY2, FY3}. Meanwhile, a lighter gravitino ($m_{3/2} < \mathcal{O}(100)$ eV) serves as warm dark matter (DM), which might be incompatible with the observed structure of the universe that strongly suggests the bottom--up type of galaxy formation. Regarding this, based on the cosmic microwave background (CMB) lensing and cosmic shear, \cite{lightgrav1} argued that $m_{3/2}$ should be smaller than $4.7$ eV. However, for such a case, the gravitino would only account for the fraction of DM relics ($\Omega_{3/2}h^2=\mathcal{O}(10^{-3})$), which may exclude the viability of gauge mediation. Moreover, the lightest Higgs boson is unlikely to receive the mass of $125$ GeV, even though the MSSM superparticle could acquire the soft mass around TeV scale;  \cite{ lightgrav2} argued that the observed Higgs mass could be embodied in the context of the vector--mediated model.

In this letter, considering a more simplified approach, we revisit the  Intriligator--Seiberg--Shih (ISS) model \cite{dual} based on the $SU(3)$ gauge group. We first note that  the Peccei--Quinn (PQ) symmetry \cite{PQ1, PQ2, PQ3, PQ4} should make the DSB mechanism most naturalized with the aid of $R$ symmetry, as an unwanted Nambu--Goldstone boson (NGB) is excluded, which eventually causes the appearance of QCD anomaly through the coupling of the DSB sector with the Higgs superfields. \cite{metastable2} already argued that $U(1)_{PQ}$, despite approximate symmetry \cite{approxiPQ, approxiPQ2, approxiPQ3}, should place strict constraints on the DSB superpotential form  so that $|\theta|<10^{-10}$ can hold ($\theta$ is the theta angle). However, the $\mu$--value is too small to be favored by the acceptable phenomenology. 

In contrast, this study shows that, via suitable $R$ charge assignment, the axion may become a good DM candidate even for $T_R>10^9$ GeV. Therein, it follows that $m_{3/2}=\mathcal{O}$ (eV) and $m_{soft}$ (the MSSM soft mass)$\simeq 1$ TeV. We also address whether the lightest Higgs boson acquires the desirable mass. \\

{\it ISS $SU(3)$ model with $N_F=4$}

We attempt to convert an NGB (as noted in \cite{M.N}) into the axion by using $U(1)_{PQ}$. Further, via the $R$ charge assignment, $m_{3/2}$ of $\mathcal{O}(eV)$ can be achieved, maintaining $f_a$ within the axion window \cite{axiwind1, axiwind2, axiwind3, axiwind4}. The $U(1)_{PQ}$ and $R$ assignment is given in Table 1, wherein an anomaly-free $Z_{8 R}$ symmetry is assumed \cite{footnote2}.
\begin{center}
\begin{tabular}{c|c}
\multicolumn{2}{c}{{\small Table 1. Charge assignment to DSB sector}}
\\ \toprule \addlinespace[2pt]
& $Q^1, Q^2, Q^3, Q^4$ \\ \hline
\addlinespace[2pt] $U(1)_{PQ}$ & $-1, \ 1/3,\  1/3,\  1/3$ \\ \hline
\addlinespace[2pt] $Z_{8 R}$ & $0,\ -1,\ -1,\ -1$  \\ \bottomrule
\end{tabular}
\end{center}
\begin{center}
\begin{tabular}{c|c}
\multicolumn{2}{c}{ }
\\ \toprule \addlinespace[2pt]
& $\bar{Q}^1, \bar{Q}^2, \bar{Q}^3, \bar{Q}^4$\\ \hline
\addlinespace[2pt] $U(1)_{PQ}$ &$1, \ -1/3,\  -1/3, \ -1/3$\\ \hline
\addlinespace[2pt] $Z_{8 R}$ & $0,\ -1, \ -1, \ -1$\\ \bottomrule
\end{tabular}
\end{center}
With a certain super Yang--Mills sector \cite{seiberg}, the superpotential of the relevance is given by
\begin{equation}
W= \dfrac{(Tr W^{\alpha}W_{\dot{\alpha}})}{M_{pl}^2} \bar{Q}^1 Q^1+  \sum_{i, j=2}^4 \dfrac{\lambda_{ij}(Tr W^{\alpha}W_{\dot{\alpha}})^2}{M_{pl}^5} \bar{Q}^iQ^j, \label{deltaQ}
\end{equation}
where $\lambda_{ij}$ is the dimensionless coupling constant. We subsequently introduce messenger superfields, $\bar{f_1}, f_1$, and $\bar{f_2}, f_2$ of $\bar{5}+5$ under $SU(5)_{GUT}$, respectively (without supposing any fundamental singlet, following \cite{M.N}). We assign the charge to each superfield in Table 2.  Accordingly, the coefficients of the QCD anomaly add up to $3$:
\begin{center}
\begin{tabular}{c|c|c|c}
\multicolumn{4}{c}{{\small Table 2.  Charge assignment to 
messenger and Higgs}}
\\ \toprule \addlinespace[2pt]
& $f_1,\  \bar{f}_1$&$f_2,\  \bar{f}_2$&$H_u,\  H_d$ \\ \hline
\addlinespace[2pt] $U(1)_{PQ}$ & $-1,\  1$ &$-1, \ 1$&$0, \ -2$\\ \hline
\addlinespace[2pt] $Z_{8 R}$ & $2,\  2$ &$-2,\  -2$&$0,\ \  0$ \\ \bottomrule
\end{tabular}
\end{center}
The relevant coupling is given by
\begin{align}
W \supset & \sum_{i, j=2}^4 \left(\gamma_{ij}\dfrac{\bar{Q}^iQ^{j}}{M_{pl}} \bar{f_1}f_1+\epsilon_{ij}\dfrac{\bar{Q}^iQ^{j}}{M_{pl}} \bar{f_2}f_2 \right)  \notag \\ &+\kappa \dfrac{(Tr W^{\alpha}W_{\dot{\alpha}})}{M_{pl}^2} \left(\bar{f_2}f_1+\bar{f_1}f_2\right) \notag \\ &+ d\dfrac{\left(Q^2Q^3 Q^4 \right)^2}{M_{pl}^5} H_u H_d,
\end{align}
with $\gamma_{ij}, \epsilon_{ij}, \kappa$, and $d$ denoting the dimensionless coupling constants as well. 

After diagonalizing the mass matrix of $\bar{Q}^i, Q^j$, the entire superpotential is then rewritten in terms of the dual picture:
\begin{align}
W_{dual}=\ &\bar{b}_iS^{ij}b_j-\dfrac{\det{S^{ij}}}{\Lambda}-m \Lambda S^{11}+ \lambda \sum_{i=2}^4\dfrac{m^2 \Lambda}{M_{pl}} S^{ii} \notag \\  &+\sum_{i, j=2}^4 \dfrac{ \Lambda}{M_{pl}} \left(\gamma^{\prime}_{ij}S^{ij} \bar{f_1}f_1+\epsilon^{\prime}_{ij}S^{ij} \bar{f_2}f_2 \right) \notag \\& + \kappa m \left(\bar{f_2}f_1+\bar{f_1}f_2\right)+d \dfrac{\Lambda^4}{M_{pl}^5} b_1^2 H_u H_d, \label{dual}
\end{align}
where $m (\equiv \Lambda^{\prime 3}/{M_{pl}^2}), \lambda \simeq \lambda_{ij}$ and $\gamma^{\prime}_{ij}, \epsilon^{\prime}_{ij}$ are expected to be of unit order. It is understood that $\lambda m^2 \Lambda/M_{pl}$ is real and positive-valued. Besides, $\Lambda^{\prime}$ is the gaugino condensation scale, and $d$ is supposed to be $\mathcal{O}(0.1-1)$ in what follows. Then, provided that $\Lambda \simeq 8 \times 10^{17}$ GeV, $m \simeq 6 \times 10^5$ GeV, and $\lambda \simeq 10^{-1.5}, \kappa \simeq 10^{-0.7}$ for instance, one obtains
\begin{gather}
m_{soft}\simeq \dfrac{1}{16 \pi^2}\Lambda_{mess}\simeq 1\ \text{TeV}, \notag \\ m_{3/2}= 3 \lambda \dfrac{m^2 \Lambda}{\sqrt{3} M_{pl}^2} \simeq 2 \ \text{eV},
\end{gather}
where 
\begin{equation} 
\Lambda_{mess}\simeq \dfrac{6 \lambda m^2 \Lambda^2}{M_{pl} \cdot m_{mess}}\simeq 10^5\ \text{GeV},
\end{equation}
and the messenger mass is given by (see Eq.(\ref{ms}))
\begin{equation}
m_{mess} = \kappa m+\dfrac{\Lambda}{M_{pl}} \left<S^{ii}\right> \simeq \kappa m.
\end{equation}
We then confirm that the flavor changing process is highly suppressed owing to $m_{soft} \gg 10^2\ m_{3/2}$. Moreover, it follows that
\begin{gather}
f_a \simeq b_1 (=\bar{b}_1) \simeq \sqrt{m \Lambda} \ \simeq 6 \times 10^{11}\ \text{GeV},\\ \left|\mu \right|=\mathcal{O}(100-1000)\ \text{GeV}.
\end{gather}
Additionally, the messenger loops induce the effective $K\ddot{a}hler$ potential of
\begin{align}
\Delta{K}\simeq &\dfrac{5}{32 \pi^2}\left(\dfrac{\Lambda}{M_{pl}}\right)^4 \notag \\ &\times \left(\dfrac{\left|\sum_{i,j=2}^4\gamma^{\prime}_{ij}S^{ij}\right|^4}{m_{mess}^2}+ \dfrac{\left|\sum_{i,j=2}^4\epsilon^{\prime}_{ij}S^{ij}\right|^4}{m_{mess}^2}\right). \label{effective-S}
\end{align}
 Consequently, the boson component of $S^{ij}$ receives a mass of
\begin{gather}
m^2_{S^{ij}} \simeq \dfrac{180 \lambda^2}{16 \pi^2 \kappa^2} \left(\dfrac{\Lambda}{M_{pl}}\right)^4 \dfrac{m^2 \Lambda^2}{M_{pl}^2}, \label{S-mass}
\end{gather}
which is much larger than the $b_i, \bar{b}_i$ loop--induced mass:  
\begin{equation}
\delta{m^2_{S^{ij}}}\simeq\dfrac{1}{16 \pi^2 m \Lambda} \left(\dfrac{\lambda m^2 \Lambda}{M_{pl}}\right)^2. \label{loop-induced-mass}
\end{equation}
 (The mass of fermion is discussed later.)

Thus, $S^{ij}$ should develop the vacuum expectation value (VEV):
\begin{align}
\left<S^{ij} \right> &\simeq \dfrac{8 \sqrt{3} \pi^2 \kappa^2}{30} \left(\dfrac{M_{pl}}{\Lambda}\right)^4 \dfrac{m^2}{M_{pl}}\notag \\ &=\mathcal{O}(10^{-6})\ \text{GeV} \   \left(\ll \dfrac{M_{pl}}{\Lambda} m \right). \label{ms}
\end{align}
through the scalar potential terms of
\begin{equation}
V\supset m^2_{S^{ij}}\left|S^{ij}\right|^2-\left(3 \lambda m_{3/2} \dfrac{m^2 \Lambda}{M_{pl}} S^{ii}+ h.c.\right),
\end{equation}
where the second term is caused by the SUGRA correction. Accordingly, via $W \supset \det{S^{ij}}/ \Lambda$, the scalar potential has the term of
\begin{align}
&V \supset \left|\dfrac{\partial W_{dual}}{\partial b_{1}}\right|^2+\left|\dfrac{\partial W_{dual}}{\partial \bar{b}_1}\right|^2+\sum_{i=2}^4\left|\dfrac{\partial W_{dual}}{\partial S^{ii}}\right|^2
\notag \\ &\supset 2 m \Lambda \left|S^{11} \right|^2-\dfrac{\lambda m^2}{M_{pl}} \left(\sum_{i, j =2}^{4}\left<S^{ii}\right> \left<S^{jj}\right> \right) S^{11}+h.c.,
\end{align} 
which slightly shifts $S^{11}$ from the origin as follows:
\begin{equation}
\left<S^{11} \right> \simeq 9 \left(\dfrac{8 \sqrt{3} \pi^2 \kappa^2}{30}\right)^2 \dfrac{\lambda m^5 M_{pl}^5}{\Lambda^9}=\mathcal{O}(10^{-40})\ \text{GeV}.
\end{equation}
Hence, it is deduced that \cite{b-mu}
\begin{gather}
\left|B_{\mu} \right|=d \dfrac{\Lambda^5 m}{M_{pl}^5} \left<S^{11}\right> \ll \left|\mu \right|^2,
\end{gather}
which implies that the electroweak symmetry remains unbroken at the messenger scale because of $m^2_{H_u}\simeq m^2_{H_d} \simeq \left|\mu \right|^2$. Besides, the stability along the $D$--flat direction can be ensured at $\left<H_u\right>=\left<H_d\right>=0$. 

Subsequently, we evaluate the axino and saxion masses. After the $U(1)_{PQ}$ breakdown, $\bar{b}_1, b_1$ are written as follows:
\begin{align}
\bar{b}_1=&\sqrt{m \Lambda}+\bar{s}_1+i\bar{a}_1+\sqrt{2}\theta \tilde{\bar{a}}_1+\theta^2F_{\bar{b}_1},\notag \\
b_1=&\sqrt{m \Lambda}+s_1+i a_1+\sqrt{2}\theta \tilde{a}_1+\theta^2F_{b_1}.
\end{align}
The tree-level scalar potential then includes the relevant terms:
\begin{equation}
V \supset \left|\dfrac{\partial W_{dual}}{\partial S^{11}}\right|^2 \supset 2 m \Lambda \left(\left|\dfrac{s_1+\bar{s}_1}{\sqrt{2}}\right|^2+\left|\dfrac{a_1+\bar{a}_1}{\sqrt{2}}\right|^2\right).
\end{equation}
Hence, the axion is (solely) identified by
\begin{equation}
a=\dfrac{a_1-\bar{a}_1}{\sqrt{2}},
\end{equation}
and accordingly, the saxion and the axino could be expressed as
\begin{equation}
s=\dfrac{s_1-\bar{s}_1}{\sqrt{2}},\ \tilde{a}=\dfrac{\tilde{a}_1-\tilde{\bar{a}}_1}{\sqrt{2}},
\end{equation}
respectively \cite{axion-multiplet}. 

Then, let us examine the fermion mass matrix of interest, which takes the form of\begin{gather}
\mathcal{M}= \begin{pmatrix}
0 & \left<S^{11}\right> & \sqrt{m \Lambda} \\
\left<S^{11}\right> & 0 & \sqrt{m \Lambda} \\
\sqrt{m \Lambda} & \sqrt{m \Lambda} & 0 
\end{pmatrix} 
\end{gather}
that leads to 
\begin{align}
 \mathcal{M^{T}} & \mathcal{M}  \notag \\  = & \begin{pmatrix}
\left<S^{11}\right>^2+m \Lambda & m \Lambda & \sqrt{m \Lambda} \left<S^{11}\right> \\ m \Lambda & \left<S^{11}\right>^2+m \Lambda & \sqrt{m \Lambda} \left<S^{11}\right> \\ \sqrt{m \Lambda} \left<S^{11}\right> & \sqrt{m \Lambda} \left<S^{11}\right> & 2 m \Lambda
\end{pmatrix}
\end{align}
The column and row run $\tilde{b}_1, \tilde{\bar{b}}_1, \tilde{S}^{11}$ (which denote the fermion component of the corresponding superfield). Hence, each mass eigenvalue and state could be given as follows:
\begin{gather}
\left<S^{11}\right>,\ \ \dfrac{\tilde{b}_1 - \tilde{\bar{b}}_1}{\sqrt{2}},\notag \\ \sim \sqrt{2 m \Lambda},\ \ \dfrac{\tilde{b}_1 + \tilde{\bar{b}}_1 \pm  \sqrt{2}\tilde{S}^{11}}{2}.
\end{gather}
Thus, the axino is found to only acquire the mass of $\mathcal{O}(10^{-40})$ GeV from the DSB sector. Meanwhile, the additional mass should be generated:
\begin{equation}
m_{\tilde{a}}\simeq \dfrac{48 \cdot 8}{16 \pi^2} \left(\dfrac{g_3^2}{64 \pi^2}\right)^2 \dfrac{M_{\tilde{g}} m_{mess}^2}{f_a^2} = \mathcal{O}(10^{-6})\ \text{eV},
\end{equation} 
due to the axino--gluino--gluon interaction of 
\begin{equation}
\mathcal{L} = i \dfrac{g_3^2}{64 \pi^2 f_a} \bar{\tilde{a}} G^a_{\mu \nu} \left[\gamma^{\mu}, \gamma^{\nu}\right] \gamma_5\tilde{g}^a,
\end{equation}where $g_3$ is the $SU(3)_{color}$ gauge coupling constant, while $M_{\tilde{g}}$ denotes the gluino mass. Eventually, we conclude that the axino has the mass of $\mathcal{O}(10^{-6})$ eV, and hence should be long--lived (or might be stable).
At this point, notice that the thermal relics should be estimated as follows:
\begin{equation}
\Omega_{\tilde{a}}h^2 \lesssim 5.8 \times 10^5 \left(\dfrac{m_{\tilde{a}}}{1\ \text{GeV}}\right)
\end{equation}
for $T_R> 10^9$ GeV. Here we considered the axino decoupling temperature of 
\begin{equation}
T_{dec.} \simeq 10^9\ \text{GeV} \left(\dfrac{f_a}{10^{11}\ \text{GeV}}\right)^2.
\end{equation}
Thus the axino abundance, even though serving as the hot DM component, is expected to be of negligible order \cite{axino-mass, axino-dens, axino1, axino2}. 

In contrast, the saxion
receives a mass of $\mathcal{O}(0.1-1)$ MeV induced by the $b_i, \bar{b}_i$ loop, as seen from Eq.(\ref{loop-induced-mass}).  
Let us estimate its life--time. Eq.(\ref{dual}) yields the relevant couplings: 
\begin{equation}
\mathcal{L} \supset \dfrac{\mu^2}{\sqrt{m \Lambda}} s \left(\left|H_u\right|^2+\left|H_d\right|^2\right) \supset \dfrac{\mu^2}{\sqrt{m \Lambda}} s \left|H^-\right|^2,
\end{equation}
\begin{equation}
\mathcal{L} \supset \dfrac{\mu}{2 \sqrt{ m \Lambda}} s\bar{\tilde{h}}_u \tilde{h}_d+ \text{h.c} =\dfrac{\mu}{2 \sqrt{ m \Lambda}} s\bar{\tilde{h}}^- \tilde{h}^-+ \text{h.c},
\end{equation}
while among the QED interaction are the following terms:
\begin{equation}
\mathcal{L}\supset \left|\left(\partial_{\mu}+i e A_{\mu}\right)H^-\right|^2,\ \bar{\tilde{h}}^-\left(\partial_{\mu}+i e A_{\mu}\right)\tilde{h}^-.
\end{equation} 
Here $H^-, \tilde{h}^-, A_{\mu}$ are the charged Higgs, higgsino and the photon with $e$ denoting the $U(1)_{em}$ coupling constant. Hence, we obtain the decay rate into two photons through $H^-$/$\tilde{h}^-$ loops:
\begin{align}
\Gamma_{s \rightarrow 2 \gamma}  \simeq \dfrac{e^4}{128 \pi^5} \dfrac{\left|\mu\right|^4}{m \Lambda m_s} \left|\mathcal{T}\right|^2\sim 10^{-16}\ \text{MeV},
\end{align}
where
\begin{align}
\mathcal{T}=&\dfrac{\Lambda_{EW}^2}{ \left|\mu \right|^2+\Lambda_{EW}^2}-\dfrac{\Lambda_{EW}^2}{m_{H^-}^2+\Lambda_{EW}^2} \notag \\ & +\log{\dfrac{m_{H^-}^2+\Lambda_{EW}^2}{m_{H^-}^2}}-\log{\dfrac{\left|\mu\right|^2+\Lambda_{EW}^2}{\left|\mu\right|^2}},
\end{align}
and $\Lambda_{EW}=\mathcal{O}(100)$ GeV is the electroweak symmetry breaking scale, while $m_s, m_W$ denote the saxion and the weak boson mass,  respectively. Hence, the saxion decay should not give rise to large  entropy production because of $T_d$ (the saxion decay temperature)$\sim 1$ GeV $\gg m_s$. To summarize, we deduce that the observed DM relics should predominantly comprise the axion as $f_a$ is close to $10^{12}$ GeV. \\

{\it The other mass spectrum}

We estimate the mass of the other particles in the DSB sector. The boson component of $S^{11}$ and $(s_1+\bar{s}_1)/\sqrt{2},\  (a_1+\bar{a}_1)/\sqrt{2}$ acquire the mass of $\sqrt{2 m \Lambda}$. Further, $S^{1i}, S^{i1}, b_i, \bar{b}_i,  (i=2-4)$  are addressed as follows. The boson components have their mass of
\begin{equation}
\sqrt{m \Lambda}\simeq 10^{12}\ \text{GeV}, 
\end{equation}
via $\left<b_1\right>=\left<\bar{b}_1 \right>=\sqrt{m \Lambda}$. On the other hand, the fermions could form the mixing mass term of
\begin{equation} 
\mathcal{L} \supset \sqrt{m \Lambda}\  \bar{\tilde{S}}_{1i} \tilde{b}_i+h.c.,\ \sqrt{m \Lambda}\  \bar{\tilde{b}}_i\tilde{S}_{i1}+h.c.,
\end{equation}
The eigenstate is then expressed as
\begin{gather}
\dfrac{\bar{\tilde{S}}_{1i} \pm \tilde{b}_i}{\sqrt{2}},\ \ \ \dfrac{\bar{\tilde{b}}_i \pm \tilde{S}_{i1}}{\sqrt{2}},
\end{gather}
all of which have the mass of $\sqrt{m \Lambda}$. 

In contrast, the boson components of $S^{jk} (j, k=2-4)$ have the mass of $m_{S^{jk}}=\mathcal{O}(10^3)$ GeV, as seen from Eq.(\ref{S-mass}) \cite{footnote-mixing}. Meanwhile, the fermion components (denoted $\tilde{S}^{jk}$) receive the mass of
\begin{align}
m_{\tilde{S}^{jk}}^{loop} & \simeq \sum_{j=2}^4 \dfrac{10}{16 \pi^2} \left(\dfrac{\Lambda}{M_{pl}}\right)^4 \left(\dfrac{\lambda  \Lambda}{\kappa^2 M_{pl}}\right)\left<S^{jj}\right> \notag \\ & = \mathcal{O}(10^{-9})\ \text{GeV}
\end{align}
from Eq.(\ref{effective-S}) (except for the goldstino approximately formed by $(\tilde{S}^{22}+\tilde{S}^{33}+\tilde{S}^{44})/\sqrt{3}$). 
Besides, an extra mass could be generated from the following coupling:
\begin{gather}
W \supset \xi_{jklmst}\left(\dfrac{\Lambda}{M_{pl}}\right)^3 S^{jk}S^{lm}S^{st},\notag \\ \ \ \ (j, k, l, m, s, t =2-4). \label{fermion-mass}
\end{gather}
Here, $\xi_{jklmst}$ (the dimensionless coupling constant) is assumed to be of unit order, for simplicity. Eventually, $\tilde{S}^{jk}$ is expected to have a mass of 
\begin{align}
m_{\tilde{S}^{jk}}=m_{\tilde{S}^{jk}}^{loop}+ \text{a few factor}\times\left(\dfrac{\Lambda}{M_{pl}}\right)^3 \left<S^{jk}\right>.
\end{align}
At this point, notice that Eq.(\ref{fermion-mass}) yields
\begin{equation}
V \supset \sum_{s=2}^4 \xi_{jklmss} \left(\dfrac{\Lambda}{M_{pl}}\right)^3 \left(\dfrac{\lambda m^2 \Lambda}{M_{pl}}\right) S^{jk}S^{lm}.
\end{equation} 
We hence require the condition that 
\begin{equation}
m^2_{S^{jk}}\gtrsim 3 \ \xi_{jklmss} \left(\dfrac{\Lambda}{M_{pl}}\right)^3 \left(\dfrac{\lambda m^2 \Lambda}{M_{pl}}\right),
\end{equation}
so as to avoid the tachyonic $S^{jk}$ bosons, which corresponds to $\xi_{jklmss} \lesssim 10^{-2}$. Therefore, it follows that 
\begin{gather}
m_{\tilde{S}^{jk}}=\mathcal{O}(10^{-7})\ \text{GeV} \ (j\neq k),\notag \\
m_{\tilde{S}^{jj}}<\mathcal{O}(10^{-9})\ \text{GeV}.
\end{gather}
 $\tilde{S}^{jk}$ is then found to decay into a photon and the goldstino (denoted $\chi$) or $\tilde{S}^{jj}$ through the messenger loop (Eq.(\ref{dual})) and the life-time is given by 
\begin{align}
\tau_{\tilde{S}^{jk}}\simeq \left(\Gamma_{\tilde{S}^{jk}\rightarrow \gamma \chi}+\Gamma_{\tilde{S}^{jk}\rightarrow \gamma \tilde{S}^{jj}}\right)^{-1}\sim 10^{-10}-10^{-9}  s
\end{align}
where each decay rate takes the form of 
\begin{align}
\Gamma_{\tilde{S}^{jk}\rightarrow \gamma \chi} \simeq \Gamma_{\tilde{S}^{jk}\rightarrow \gamma \tilde{S}^{jj}} =\dfrac{m_{\tilde{S}^{jk}}}{16 \pi} \left|\mathcal{M}\right|^2
\end{align}
with 
\begin{align}
\mathcal{M} \simeq\dfrac{e}{8 \pi^2} \left(\dfrac{\Lambda}{M_{pl}}\right)^2 \left(1+\mathcal{O}\left( \dfrac{\Lambda}{M_{pl}} \dfrac{F_{S^{jj}}}{m^2_{mess}}\right)^2\right),
\end{align}
where $F_{S^{jj}}=\lambda m^2 \Lambda/M_{pl}$. (the goldstino is identified by the longitudinal mode of the gravitino.)  
Consequently, the relic abundance of the gravitino or $\tilde{S}^{jj}$ is estimated as 
\begin{gather}
\Omega_{3/2}h^2=\Omega_{3/2}^{TH}h^2+\Omega_{3/2}^{NTH}h^2=\mathcal{O}(10^{-3}),\notag \\
\Omega_{\tilde{S}^{jj}} h^2=\Omega_{\tilde{S}^{jj}}^{TH}h^2+\Omega_{\tilde{S}^{jj}}^{NTH}h^2=\mathcal{O}(10^{-3}),
\end{gather}
both of which have less cosmological implications. Here, $\Omega^{TH}h^2, \Omega^{NTH}h^2$ denote thermally and nonthermally produced relics, respectively, and we suppose that the number density of $\tilde{S}^{jk}$ to entropy is given by
\begin{equation}
\dfrac{n_{\tilde{S}^{jk}}}{s}\simeq 10^{-3}
\end{equation}
at the reheating epoch \cite{nonthermal-grav-relic}. \\

{\it Enhancing the lightest Higgs mass}

We briefly address the enhancement of the lightest Higgs boson mass. With the stop mass of around $10$ TeV, the observed mass of $125$ GeV \cite{atlas, cms} would be apparently embodied; a class of gauge mediation model with $m_{3/2}$ of $\mathcal{O}(eV)$ generally indicates $m_{soft}$ less than a few TeV. We could, however, consider an alternative source via another pair of messenger:
\begin{align}
\Delta{W}= &  \ c \  \dfrac{\Lambda^4 \bar{b}_1 b_1}{M_{pl}^5} \bar{f}_3 f _3 \notag \\&+ \sum_{j, k=2}^4 h_{jk} \dfrac{\Lambda}{M_{pl}} H_u \bar{f}_3 S^{jk},\  (j, k=2-4) \label{higgs-mass-coupling}
\end{align}  
where $\bar{f}_3, f_3$ belong to $\bar{5}+5$ of $SU(5)_{GUT}$ which are assumed to carry the $U(1)_{PQ}$ and $Z_{8 R}$ charge of $(0, 0)$,  $(4, 4)$, respectively. (Note that our analysis still avoids the Landau pole up to GUT scale.)  $c, h_{jk}$ are the dimensionless coupling constants. Through $S^{jk}, \bar{f}_3$ and $f_3$ loops, the extra scalar potential should be generated (the other source is omitted because it yields only a small contribution):
\begin{align}
\Delta{V} &\simeq \sum_{j, k=2}^4 \dfrac{1}{16 \pi^2} \left(M_{b +}^4 \log{\dfrac{M_{b +}^2}{\Lambda_{cut}^2}}+M_{b -}^4 \log{\dfrac{M_{b -}^2}{\Lambda_{cut}^2}}\right)\notag \\&-\sum_{j, k=2}^4 \dfrac{1}{16 \pi^2} \left(M_{f +}^4 \log{\dfrac{M_{f +}^2}{\Lambda_{cut}^2}}+M_{f -}^4 \log{\dfrac{M_{f -}^2}{\Lambda_{cut}^2}}\right), 
\end{align}
where 
\begin{align}
&M_{b \pm}^2=\dfrac{1}{2}\left(m^2_1+\left|h_{jk}\right|^2 \dfrac{\Lambda^2 }{M_{pl}^2}\left|H_u \right|^2\right)\notag \\& \pm \dfrac{1}{2}\sqrt{\left(m_1^2+\left|h_{jk}\right|^2 \dfrac{\Lambda^2 }{M_{pl}^2}\left|H_u \right|^2\right)^2-4 m^{\prime 2} m_{S^{jk}}^{ 2}},\\ &
M_{f \pm}^2=\dfrac{1}{2}\left(m^2_2+\left|h_{jk}\right|^2 \dfrac{\Lambda^2 }{M_{pl}^2}\left|H_u \right|^2\right)\notag \\& \pm \dfrac{1}{2}\sqrt{\left(m_2^2+\left|h_{jk}\right|^2 \dfrac{\Lambda^2 }{M_{pl}^2}\left|H_u \right|^2\right)^2-4 m^{\prime 2} m_{\tilde{S}^{jk}}^{ 2}},\label{higgsmass-loop}
\end{align}
with $m^{\prime} \equiv \left|c \right| m \Lambda^5 /M_{pl}^5$ and $m_1^2\equiv m^{\prime 2}+m_{S^{jk}}^{ 2},\  m_2^2\equiv m^{\prime 2}+m_{\tilde{S}^{jk}}^{ 2}$. Besides, $\Lambda_{cut}$ denotes some cut off scale (which has no significant effect on the following discussion).  
Postulated that
\begin{equation}
m^{\prime} \lesssim \left|h_{jk}\right| \dfrac{\Lambda }{M_{pl}}\left<H_u \right>,
\end{equation}
the quartic Higgs coupling should be approximately reduced to the form of
\begin{align}
\Delta{V}&\simeq \sum_{j, k=2}^4 \dfrac{\left|h_{jk}\right|^4}{32 \pi^2} \left(\dfrac{\Lambda}{M_{pl}}\right)^4 \left|H_u \right|^4 \notag \\  & \times \log{\left( \dfrac{m_{S^{jk}}^{ 2}+\left|h_{jk}\right|^2 \dfrac{\Lambda^2 }{M_{pl}^2}\left|H_u \right|^2}{\left|h_{jk}\right|^2 \dfrac{\Lambda^2 }{M_{pl}^2}\left|H_u \right|^2}\right)}, \label{quartic-coupling}
\end{align}
through the relation that
\begin{equation}
m_{S^{jk}} \gg \left|h_{jk}\right| \dfrac{\Lambda }{M_{pl}}\left<H_u \right>,
\end{equation}
where $\left<H_u \right>=v \sin{\beta}$ and $v \simeq 246$ GeV is the Higgs VEV. Incorporating the contributions from Eq.(\ref{quartic-coupling}) and the stop/top quark loop, the lightest Higgs boson receives the mass of
\begin{align}
 m_{H^0}^2 & \simeq  m^2_Z \cos^2{2 \beta}+\dfrac{3}{4 \pi^2} \dfrac{m^4_t}{v^2} \log{\left(\dfrac{m^2_{stop}}{m^2_t}\right)} \notag \\  + &  \sum_{j, k=2}^4 \dfrac{\left|h_{jk}\right|^4 v^2}{8 \pi^2} \left(\dfrac{\Lambda}{M_{pl}}\right)^4  \sin^4{\beta} \notag \\   \times &
\log{\left(\dfrac{m_{S^{jk}}^{ 2}+\left|h_{jk}\right|^2\dfrac{\Lambda^2 }{M_{pl}^2}v^2 \sin^2{\beta}}{\left|h_{jk}\right|^2\dfrac{\Lambda^2 }{M_{pl}^2}v^2 \sin^2{\beta}}\right)}.
\end{align}
 At this stage, noting that Eq.(\ref{higgs-mass-coupling}) yields
\begin{equation}
V \supset \sum_{j=2}^4 h_{jj} \dfrac{\Lambda}{M_{pl}} \dfrac{\lambda m^2 \Lambda}{M_{pl}} H_u \bar{f}_3+h.c.,
\end{equation}
we require the following condition:
\begin{equation}
m^{\prime} \cdot \dfrac{m \Lambda^5}{M_{pl}^5} \gtrsim  \sum_{j=2}^4 \left|h_{jj}\right| \dfrac{\Lambda}{M_{pl}} \dfrac{\lambda m^2 \Lambda}{M_{pl}},
\end{equation}
so that $\bar{f}_3$ and the charged Higgs boson cannot develop their VEVs.  Hence, given $\left|c \right| = \mathcal{O}(10^{-2})$ for instance, $\left|h_{jj}\right| < \mathcal{O}(10^{-5})$ should be entailed. Taking account of the other radiative correction \cite{higgs-mass-correction, higgs-mass-correction2, higgs-mass-correction3}, we thus obtain $m_{H^0}\simeq 125$ GeV for $\left|h_{jk} \right| \sim1.6-1.7 \ (j \neq k), \ \tan{\beta} \sim 10$. \\

{\it Discussion}

Let us discuss the (meta) stability of vacuum.
The longevity of the SUSY breaking vacuum should be ensured owing to the relation of
\begin{equation}
 m, \ \dfrac{m^2}{M_{pl}} \ll \Lambda.
\end{equation}
 Moreover, it is verified that
\begin{equation}
m_{mess}^2> \dfrac{3 \lambda m^2 \Lambda}{M_{pl}}, \label{no-tachyon}
\end{equation}
which should avoid the tachyonic messengers.
Accordingly, there is no local minimum state along the messenger direction \cite{footnote4}. 

Further, the PQ symmetry should not be restored after the inflationary epoch, as long as it is expected that
\begin{equation}
f_a> T_{max} = \mathcal{O}(10) \cdot T_R,
\end{equation}
for $T_R = \mathcal{O}(10^9-10^{10})$ GeV, and consequently, the domain wall (DW) problem may not be encountered in our framework. Here $T_{max}$ is the maximal temperature at the reheating stage. \\

{\it Conclusion}

We argued that the ISS gauge mediation could avoid the gravitino problem, given the PQ charged DSB sector. Thereby, we first proposed that $(b_1-\bar{b}_1)/\sqrt{2}$ could be identified by the axion supermultiplet.  This naturally generates the QCD anomaly and a sizable $\mu$--value. Then, we eventually confirmed that the axion should predominantly account for the DM density instead of the light gravitino. Further, the lightest Higgs boson is likely to acquire the desirable mass, owing to several dual singlets and an additional pair of messenger.

Finally, we have not addressed the Izawa--Yanagida--Intriligator--Thomas (IYIT)  DSB mechanism \cite {DSB, DSB2} based on the $SU(3)$ gauge group with $N_F=3$ \cite{ponton}. Therein, the PQ symmetry could appropriately constrain the DSB superpotential form, such that unwanted NGBs should be absent \cite{metastable2}. This model might provide  $m_{3/2}=\mathcal{O}(1)$ eV and $m_{soft} \simeq 1$ TeV for $f_a=\mathcal{O}(10^9)$ GeV, whereas the DW could appear. Under such circumstances, the DW decay, possibly entailed by the approximate PQ symmetry, produces the axion, which may almost saturate the observed DM density \cite{sikivie, k-h}, although the appearance of appropriate $U(1)_{PQ}$ violating should be discussed in future research.  \\

{\section* {Acknowledgment}}
{\flushleft \ }
I would like to thank Prof. Yanagida (Kavli IPMU (WPI), UTIAS, The University of Tokyo, and T. D. Lee Institute and School of Physics and Astronomy, Shanghai Jiao Tang University) for motivating the concept of this study at its early stage.\\


\begin{thebibliography}{99}
\bibitem{footnote2}
{\small At this point, we state that no $Z_{8 R} [SU(3)]^2$ anomalies appear. Moreover, $Z_{2 R}$ or $Z_{4 R}$ symmetry, which is also anomaly--free, would not forbid the bare mass terms of $\bar{Q^i}, Q^i$.}
\bibitem{b-mu}
{\small At low--energy scale, the renormalization group flow should enhance $|B_{\mu}/\mu|$ up to the magnitude of several hundred GeV, i.e., $\mathcal{O}(0.1)\times M_{\tilde{w}}$ ($M_{\tilde{w}}$ is the wino mass) \cite{RG, RG2}. We, thus, expect that with $\left|\mu\right|=\mathcal{O}(100)$ GeV for $d=\mathcal{O}(0.1)$, the electroweak symmetry could break down at weak scale.}
\bibitem{axion-multiplet}
\small{Hence, we conclude that the axion supermultiplet (denoted $A$) is approximately expressed as
\begin{equation}
\dfrac{b_1-\bar{b}_1}{\sqrt{2}}=\sqrt{m \Lambda} \exp{\left(\dfrac{A}{\sqrt{m \Lambda}}\right)}.
\end{equation}}
\bibitem{footnote-mixing}
{\small Actually, the boson (and the fermion as mentioned below) should be described as each mass eigen state.}
\bibitem{nonthermal-grav-relic}
\small{The nonthermally produced gravitino relics may be cosmologically harmless, because it is expected that its free streaming length should be much smaller than $\mathcal{O}(10^{-1})$ Mpc (See Appendix C in \cite{axino1}).}
\bibitem{footnote4}
{\small Given some lower-energy state depending on $\lambda_{ij}, \kappa$, its longevity can be ensured owing to
\begin{equation}
\lambda \kappa m^3 \gtrsim \left(\dfrac{\lambda m^2 \Lambda}{M_{pl}}\right)^{3/2}.
\end{equation}}
\end{thebibliography}
\end{document}